\documentclass[10pt,twoside]{article}
\usepackage{Latex-document}
\almvol{00}
\almttone{Frontiers of Science Awards}
\almtttwo{for Math/TCIS/Phys}
\firstpage{1}
\usepackage[english]{babel}
\usepackage[utf8]{inputenc}
\usepackage[numbers,sort&compress]{natbib}

\usepackage[T1]{fontenc}
\usepackage{lmodern}
\usepackage{microtype}

\newcommand{\Gtwo}{G_2{}}
\newcommand{\Gthree}{G_3{}}
\newcommand{\Gfour}{G_4{}}
\newcommand{\Gfive}{G_5{}}

\newcommand{\Hfour}{F_4{}}
\newcommand{\Hfive}{F_5{}}

\usepackage{multicol}

\usepackage{amsmath}
\usepackage{amssymb}
\usepackage{graphicx}
\usepackage{calc}

\numberwithin{equation}{section}

\DeclareRobustCommand{\okina}{%
  \raisebox{\dimexpr\fontcharht\font`A-\height}{%
    \scalebox{0.8}{`}%
  }%
}

\begin{document}

\markboth{\hfill{\rm Jeremy Sakstein and Bhuvnesh Jain} \hfill}{\hfill {\rm The Speed of Gravity and the Fate of Dark Energy \hfill}}

\title{The Speed of Gravity and the Fate of Dark Energy}
\author{
Jeremy Sakstein$^1$ and Bhuvnesh Jain$^2$\\
{\small $^1$Department of Physics \& Astronomy, University of Hawai\okina i, Honolulu, HI 96822, USA}\\
{\small \texttt{sakstein@hawaii.edu}}\\
{\small $^2$Department of Physics and Astronomy, Center for Particle Cosmology, University of Pennsylvania, 209 S.~33rd St., Philadelphia, PA 19104, USA}\\
{\small \texttt{bjain@physics.upenn.edu}}
}

\begin{abstract}
On August 17$^{\rm th}$ 2017, observatories worldwide made a landmark detection: gravitational waves and light from a binary neutron star merger. This event revolutionized our understanding of astrophysics, cosmology, and gravitation. In this proceeding of the 2025 International Congress of Basic Science, we describe how it transformed our view of cosmic acceleration (dark energy). The near-simultaneous arrival of light and gravitational waves shows that their speeds agree to within one part in $10^{15}$, excluding large classes of modified gravity theories and interactions between dark energy and matter.
\end{abstract}

\maketitle



\setcounter{tocdepth}{1}
\tableofcontents

\section{Introduction}

In 1998, astronomers made a surprising discovery:~the cosmic expansion is accelerating \cite{SupernovaSearchTeam:1998fmf,SupernovaCosmologyProject:1998vns}.~We expect  the Universe to be dominated by gravity on its largest scales, but this is an attractive force so a different kind of physics is needed to understand what is pushing the cosmos apart.~Today, more than 25 years later, we still don't know the physics underlying the cosmic acceleration, but we've given it a name:~\textit{Dark Energy}.~Discovering the microphysical origin of dark energy remains a paramount goal of modern cosmology.

John Wheeler --- one of the pioneers of general relativity --- famously said:
\begin{quote}
   \textit{Spacetime tells matter how to move;~matter tells spacetime how to curve.}
\end{quote}
This provides a guide to what dark energy could possibly be.~Either there's a new form of matter that pushes the Universe apart (matter tells spacetime how to curve);~or Einstein's theory of general relativity is modified on large scales (spacetime tells matter how to move).

The simplest form of new matter one can consider is a cosmological constant $\Lambda$ --- a space-time filling fluid with negative pressure that drives the Universe to accelerate.~Until recently, this was our best-fitting model of dark energy \cite{Planck:2015fie,DES:2021wwk,DESI:2025zgx} but it suffers from theoretical fine-tuning issues \cite{Burgess:2013ara,Padilla:2015aaa} which motivated a search for alternatives over the past 25 years.~Models beyond the cosmological constant introduce more generalized fluids driven by new particles such as scalars and vectors \cite{Copeland:2006wr,Bull:2015stt,Bahamonde:2017ize}.~They differ from a cosmological constant in their prediction that equation of state of dark energy, $w=P/\rho$, is time-varying.

The competing hypothesis --- that gravity is modified --- is more difficult to realize.~General relativity has been tested in a variety of systems on scales ranging from laboratories to the solar system to compact objects for over 100 years \cite{Murata:2014nra,Will:2014kxa,Burrage:2016bwy,Burrage:2017qrf,Sakstein:2017pqi,Sakstein:2018fwz,Baker:2019gxo,Brax:2021wcv,Kramer:2021jcw,LIGOScientific:2021sio,Yunes:2024lzm}.~In general, any new force from modified gravity is constrained to be at least 1000 times weaker than general relativity.~It is difficult to imagine how such a weak force could come to dominate the Universe and, indeed, many modified gravity models are not viable for this reason.

A major breakthrough came in 2003 with the discovery of \textit{screening mechanisms} \cite{Khoury:2003aq,Khoury:2003rn}.~Screening mechanisms introduce nonlinearities into modified gravity theories that result in the modifications becoming environmentally-dependent \cite{Joyce:2014kja,Burrage:2016bwy,Burrage:2017qrf,Sakstein:2018fwz,Brax:2021wcv}.~Modified gravity effects are weakened in high density environments where we have tested general relativity but become important on cosmological scales.~This enables them to drive the cosmic acceleration while remaining compatible with tests of general relativity.

The \textit{Vainshtein mechanism} \cite{Vainshtein:1972sx} quickly emerged as a promising foundation for building modified gravity models of dark energy.~Theories such as galileons \cite{Nicolis:2008in} and massive gravity \cite{deRham:2010kj,deRham:2014zqa} predicted self-accelerating Universes (without a cosmological constant or new fluid) and made predictions that agreed with general relativity on smaller scales.~These successes motivated two questions\footnote{In what follows, \textit{most general} is taken to mean that the theory is free of ghost pathologies.}:
\begin{enumerate}
    \item What is the most general coupling that a dark energy scalar can have to matter? and
    \item What is the most general scalar-tensor theory that exhibits the Vainshtein mechanism?
\end{enumerate}
The answer to question one was quickly found to be conformal-disformal couplings via a Jordan frame metric ($\tilde{g}_{\mu\nu}$) of the form:
\begin{equation}
\label{eq:disformal}\tilde{g}_{\mu\nu}=A(\phi,X)g_{\mu\nu}+B(\phi,X)\partial_\mu\phi\partial_\nu\phi,
\end{equation}
where $\phi$ is the scalar, $X=-(\partial_\mu\phi\partial^\mu\phi)/2$, $g_{\mu\nu}$ is the Einstein frame metric and $A$ and $B$ are arbitrary functions.~It turns out that the answer to question 2 was partially answered in 1974 by Horndeski \cite{Horndeski:1974wa}.~The \textit{Horndeski theory} was extended in 2014 by Gleyzes, Langlois, Piazza, and Vernizzi (GLPV) to arrive at the most general model:~\textit{beyond Horndeski} or \textit{GLPV} \cite{Gleyzes:2014dya,Gleyzes:2014qga}.~

Together with the couplings in equation~\eqref{eq:disformal}, the GLPV theory formed the foundation for building modified gravity models of dark energy that were consistent with tests of general relativity.~They were excluded in 2017 by a landmark event.~On August 17$^{\rm th}$ 2017, the LIGO/Virgo gravitational wave interferometers detected gravitational waves from a binary neutron star merger --- GW170817.~Shortly thereafter, the Fermi  telescope observed a gamma ray burst --- GRB170817A.~The two events were quickly determined to have originated from the same source --- a binary neutron star merger in the galaxy NGC 4993 \cite{LIGOScientific:2017vwq}.~In this work, we review the implications of this event for cosmological scalar-tensor theories of dark energy, originally described in our Frontiers of Science Award-winning paper \cite{Sakstein:2017xjx}.

We begin with the necessary background for interpreting GW170817/GRB170817A in the GLPV framework.~Next, we describe the three results of \cite{Sakstein:2017xjx}:
\begin{enumerate}
    \item Quartic galileon models --- the quintessential paradigm for Vainshtein screening ---  are excluded as dark energy.
    \item GLPV theories  can be directly constrained by astrophysical tests.
    \item The disformal coupling of dark energy to photons is tightly constrained.
\end{enumerate}
We conclude by reviewing the broader implications of GW170817/GRB170817A for dark energy model building and summarizing the state of dark energy in 2025.

\textbf{Conventions:}~We work in units where $\hbar=c=1$ and use the $(-,+,+,+)$ metric signature convention.~The Planck mass is $M_{\rm Pl}=1/\sqrt{8\pi G}$.~Cosmological space-times are taken to be of the flat Friedmann-Lema\^{i}tre-Robertson-Walker (FLRW) class, which, in coordinate time, is
\begin{equation}
    \label{eq:FLRW}
    g_{\mu\nu}=\textrm{diag}(-1,a(t),a(t),a(t)),
\end{equation}
with $a(t)$ the cosmological scale factor and where the spatial coordinates are Cartesian $x,y,z$.

\section{GLPV Theories}

In this section we review the salient features of GLPV theories.~The reader is referred to reviews e.g., \cite{Kobayashi:2019hrl} , and to original papers therein for further details.

\subsection{Action}

The action for GLPV theories is given by 
\begin{equation}
    S=\int {\mathrm{d}}^4x\sqrt{-g}\sum_{i=2}^5L_i
    \label{eq:fullaction}
\end{equation}
where
\begin{align}
L_2 & =  \Gtwo(\phi,X)\;,  \label{L2} \\
L_3 & =  \Gthree(\phi, X) \, \Box \phi \;, \label{L3} \\
L_4 & = \Gfour(\phi,X) \, R + \Gfour_{,X}(\phi,X) ((\Box \phi)^2 - \phi^{ \mu \nu} \phi_{ \mu \nu}) \nonumber \\
&+\Hfour(\phi,X) \epsilon^{\mu\nu\rho}_{\ \ \ \ \sigma}\, \epsilon^{\mu'\nu'\rho'\sigma}\phi_{\mu}\phi_{\mu'}\phi_{\nu\nu'}\phi_{\rho\rho'}\;, \label{L4} \\
L_5 & = \Gfive(\phi,X) \, \!G_{\mu \nu} \phi^{\mu \nu}  \nonumber \\
&-  \frac16  \Gfive_{,X} (\phi,X) ((\Box \phi)^3 - 3 \, \Box \phi \, \phi_{\mu \nu}\phi^{\mu \nu} + 2 \, \phi_{\mu \nu}  \phi^{\mu \sigma} \phi^{\nu}_{\  \sigma}) \nonumber \\ \;&
+\Hfive (\phi,X) \epsilon^{\mu\nu\rho\sigma}\epsilon^{\mu'\nu'\rho'\sigma'}\phi_{\mu}\phi_{\mu'}\phi_{\nu \nu'}\phi_{\rho\rho'}\phi_{\sigma\sigma'} \label{L5}\,,
\end{align}
where $L_i$ and $F_i$ are arbitrary functions.~The theory is free of ghost pathologies provided that one of $L_{i}$ or $F_{i}$ is zero with $i=3,4$ i.e., one cannot have both an $F_i$ and an $L_i$ term simultaneously \cite{Crisostomi:2016tcp}.

\subsection{The Vainshtein Mechanism}
\label{sec:VainshteinMechanism}

As noted above, the action in Eq.~\eqref{eq:fullaction} gives rise the Vainshtein screening mechanism in all but the most trivial cases.~To exemplify this, we will consider the \textit{quartic galileon} model \cite{Deffayet:2009wt}, a paragon of GLPV dark energy theories.~This model has
\begin{equation}
    \label{eq:quarticGalileon}
G_2(\phi,X)=X;\quad\textrm{and}\quad G_4(\phi,X)=\frac{{M_{\rm Pl}}^2}{2}\left(1+2c_0\frac{\phi}{{M_{\rm Pl}}}\right)+2\frac{c_4}{\Lambda_4^6}X^2,
\end{equation}
where $\Lambda_4$ is a new mass scale.~For a static spherically symmetric object with mass $M$ and density $\rho(r)$ e.g., the Sun, the equation of motion for the field is
\begin{equation}
    \label{eq:QuarticGalEOMSun}
   \frac{\mathrm{d}}{\mathrm{d}r}(r^2\phi')+\frac{4c_4}{\Lambda_4^6}\frac{\mathrm{d}}{\mathrm{d}r}(\phi')^3 = c_0\frac{r^2\rho(r)}{M_{\rm Pl}},
\end{equation}
where $'=\mathrm{d}/\mathrm{d}r$.~This equation can be solved in two regimes.~When the second term on the left hand side is negligible one finds that 
\begin{equation}
    \phi'=c_0\frac{M}{4\pi M_{\rm Pl}r^2}
\end{equation}
while when the first term is negligible one has
\begin{equation}
    \phi'=\frac{\Lambda_4^2}{2}\left(\frac{c_0M}{c_4M_{\rm Pl}}\right)^{\frac13}.
\end{equation}
Setting these solutions equal, one find that the crossover occurs at a \textit{Vainshtein radius}
\begin{equation}
    r_V=\frac{c_4^{\frac16}}{\Lambda_4}\left(\frac{c_0 M}{2\pi M_{\rm Pl}}\right)^{\frac13}.
    \label{eq:VainshteinRadius}
\end{equation}
The first solution is valid when $r\gg r_V$ while the second is valid when $r\ll r_V$. The ratio of the Newtonian force $F_N$ to the scalar-mediated fifth force --- given by $F_\phi=c_0 \phi'/M_{\rm Pl}$ \cite{Sakstein:2013pda,Sakstein:2014jrq}  --- is:
\begin{equation}
    \frac{F_\phi}{F_N}=2c_0^2
    \begin{cases}
    \left(\frac{r}{r_V}\right)^2, & r<r_V \\
    1, & r>r_V
\end{cases}.
\end{equation}
This is the Vainshtein mechanism:~when $r<r_V$ the scalar force is suppressed by a factor of $(r/r_V)^2$ compared with the Newtonian force but when $r>r_V$ the two forces are comparable.~For typical dark energy theories, $r_V\sim\mathcal{O}(100\, {\rm pc})$ \cite{Sakstein:2018fwz,Baker:2019gxo,Brax:2021wcv}, so the Sun Vainshtein screens the entire solar system.

\subsection{Dark Energy and the Speed of Gravity in GLPV Theories}

While the free functions in equations \eqref{L2}--\eqref{L5} provide a large amount of freedom in constructing GLPV theories, the background cosmology and dynamics of linear cosmological perturbations are governed by six free functions of cosmic time $t$ \cite{Bellini:2014fua,Gleyzes:2014qga}:
\begin{align}
    M^2\,\,&\textrm{--- the time-dependent Planck mass} \\
    \alpha_M\,\, &\textrm{--- rate of change in the Planck mass}\\
    \alpha_K\,\, &\textrm{---  kinetic energy of scalar perturbations}\\
    \alpha_B\,\, &\textrm{--- amount of scalar-graviton kinetic mixing}\\
     \alpha_T=c_T^2-1\,\, &\textrm{--- tensor speed excess}\\
     \alpha_H\,\, &\textrm{--- deviation from Horndeski theories}.
\end{align}
Collectively, these are known as the \textit{building blocks of dark energy}.~When building GLPV theories of dark energy, one can choose functional forms for these six quantities instead of specifying the free functions in \eqref{L2}--\eqref{L5}.~

In the context of the present work, the functions $\alpha_B$, $\alpha_T$, and $\alpha_H$ are most relevant.~Further details on the others can be found in \cite{Bellini:2014fua}.~The \textit{braiding} function $\alpha_B$ parameterizes the degree with which the scalar and graviton are kinetically mixed e.g., via terms in the Lagrangian of the form $R_{\mu\nu}\partial^\mu\phi\partial^\nu\phi$.~When such terms are present, linear cosmological perturbations differ from those of  perfect fluid dark energy such as quintessence and K-essence.~The \textit{beyond Horndeski} function $\alpha_H$ parameterizes the new contributions of GLPV theories that are not present in the Horndeski gravity class of models;~it is present whenever $F_4$ or $F_5$ are non-zero.~Finally, $\alpha_T$ parameterizes the speed of tensor perturbations --- gravitational waves.

In general relativity, the speed of gravitational waves is identical to the speed of light, but in GLPV theories, the two speeds differ on space-times which break Lorentz invariance, including the FLRW metric.~On such metrics, the speed of gravitational waves is denoted as $c_T^2$ while the speed of light is simply unity (or $c$ if one is not working in natural units).~GLPV theories that can account for dark energy typically predict that the speed of gravitational waves and the speed of light differs appreciably ($|c_T^2-1|=|\alpha_T|\gg1$) on FLRW backgrounds \cite{DeFelice:2011bh,Bellini:2014fua,Lombriser:2015sxa,Bettoni:2016mij}.~As described below, it is this prediction that ultimately sealed the fate of GLPV dark energy theories.

\subsection{Vainshtein Breaking}

When either of the $F_4$ or $F_5$ terms in equation~\eqref{eq:fullaction} are non-zero, GLPV theories exhibit a phenomenon called \textit{Vainshtein breaking} whereby the Vainshtein mechanism is effective at suppressing fifth-forces outside extended objects but potentially large deviations from Newtonian gravity occur inside \cite{Kobayashi:2014ida,Koyama:2015oma,Saito:2015fza}.~For a spherically-symmetric body, the equations for the Newtonian potential $\Phi$ and the lensing potential $\Psi$ are modified as
\begin{align}
    \label{eq:VainshteinBreaking1}
    \frac{\mathrm{d}\Phi}{\mathrm{d}r} &=\frac{G_N M(r)}{r^2}+\Upsilon_1\frac{G_NM''(r)}{4}\quad\textrm{and}\\\frac{\mathrm{d}\Psi}{\mathrm{d}r} &=\frac{G_N M(r)}{r^2}-\Upsilon_2\frac{5G_NM'(r)}{4r}\label{eq:VainshteinBreakin21},
\end{align}
where the metric is
\begin{equation}
    \mathrm{d} s^2 =-(1+2\Phi(r))\mathrm{d} t^2 + (1-2\Psi(r))\delta_{ij}\mathrm{d}{x}^i\mathrm{d}x^j.
\end{equation}
The parameters $\Upsilon_i$ are  functions of the free functions appearing in the GLPV theory evaluated on the cosmological background \cite{Saito:2015fza,Langlois:2017dyl,Dima:2017pwp,Crisostomi:2017lbg}.~Of particular interest for this proceeding is the case of quartic GLPV theories, where $\Upsilon_i$ are directly related to the building blocks of dark energy via:
\begin{align}
\Upsilon_1&=\frac{4\alpha_H^2}{(1+\alpha_T)(1+\alpha_B)-\alpha_H-1}\quad\textrm{and}\label{eq:U1}\\
\Upsilon_2&=\frac{4\alpha_H(\alpha_H-\alpha_B)}{5[(1+\alpha_T)(1+\alpha_B)-\alpha_H-1]}.\label{eq:U2}
\end{align}
Vainshtein breaking enabled GLPV theories of dark energy to be tested in a variety of astrophysical bodies \cite{Koyama:2015oma,Saito:2015fza,Sakstein:2015zoa,Sakstein:2015aac,Sakstein:2015aqx,Jain:2015edg,Sakstein:2016ggl,Babichev:2016jom,Sakstein:2016lyj,Sakstein:2016oel,Salzano:2017qac,Saltas:2018mxc,Saltas:2019ius,Saltas:2022ybg}.

\section{GW170817/GRB170817A and their Implications for GLPV Theories}

As noted above, the LIGO  interferometers detected gravitational waves from coalescing binary neutron stars on August 17th 2017 --- an event dubbed GW170817 \cite{LIGOScientific:2017vwq}.~An optical counterpart, the gamma-ray burst GRB170817A was observed by the Fermi gamma-ray telescope \cite{LIGOScientific:2017zic} and several optical telescopes \cite{LIGOScientific:2017ync} 1.7s after this detection.

Provided the distance to the event is known this observation can be used to test the prediction of GLPV dark energy theories --- described above --- that the speed of light and gravitational waves differ appreciably.~The Virgo detector was operational during the event but did not detect a signal, indicating that the merger occurred in its blind spot.~This knowledge aided in the search effort, and the event was ultimately localized to NGC 4993, a galaxy 40 Mpc away from the Milky Way.

The observation that light and gravitational waves arrived within 1.7s of each other provides a new and stringent bound on the difference between their speeds \cite{Sakstein:2017xjx}:
\begin{equation}
\label{eq:speedBound}
    c^2_{\rm light}-c^2_{\rm GW}=c_T^2-1=\alpha_T<6\times10^{-15}.
\end{equation}
Such a negligible difference has strong implications for GLPV theories and scalar field models of dark energy more generally.~We now describe the three of these presented by \cite{Sakstein:2017xjx}.

\subsection{The building blocks of quartic GLPV dark energy theories can be constrained directly using astrophysical bodies}

As shown in equations \eqref{eq:U1} and \eqref{eq:U2}, the deviations from Newtonian gravity inside astrophysical bodies --- equations \eqref{eq:VainshteinBreaking1} and   \eqref{eq:VainshteinBreakin21} --- depend on three of the building blocks of dark energy --- $\alpha_H$, $\alpha_B$, and $\alpha_T$ so bounds on $\Upsilon_i$ give degeneracies between bounds on these three $\alpha$ functions.~The new constraint from GW170817/GRB170817A sets $\alpha_T\ll1$.~Meanwhile, bounds on $\Upsilon_i$, and accordingly $\alpha_B$, $\alpha_H$, and $\alpha_T$ from astrophysical probes were, at the time, at the $10^{-2}$ level at best \cite{Sakstein:2015zoa}.~This implies that one can set $\alpha_T=0$ in equations \eqref{eq:U1} and \eqref{eq:U2} for all intents and purposes, in which case $\Upsilon_i$ only depends on two functions $\alpha_B$ and $\alpha_H$:
\begin{align}
\Upsilon_1&=\frac{4\alpha_H^2}{\alpha_B-\alpha_H}\quad\textrm{and}\label{eq:U1NEW}\\
\Upsilon_2&=-\frac{4}{5}\alpha_H.\label{eq:U2NEW}
\end{align}
With two $\Upsilon_i$ parameters now depending on only two building blocks of dark energy, the previous degeneracy is broken and astrophysical bounds on $\Upsilon_i$ can be directly translated into bounds on the cosmology of GLPV theories.~This important result connects cosmology with small-scale tests of gravity.

\begin{figure}
    \centering
\includegraphics[width=0.5\linewidth]{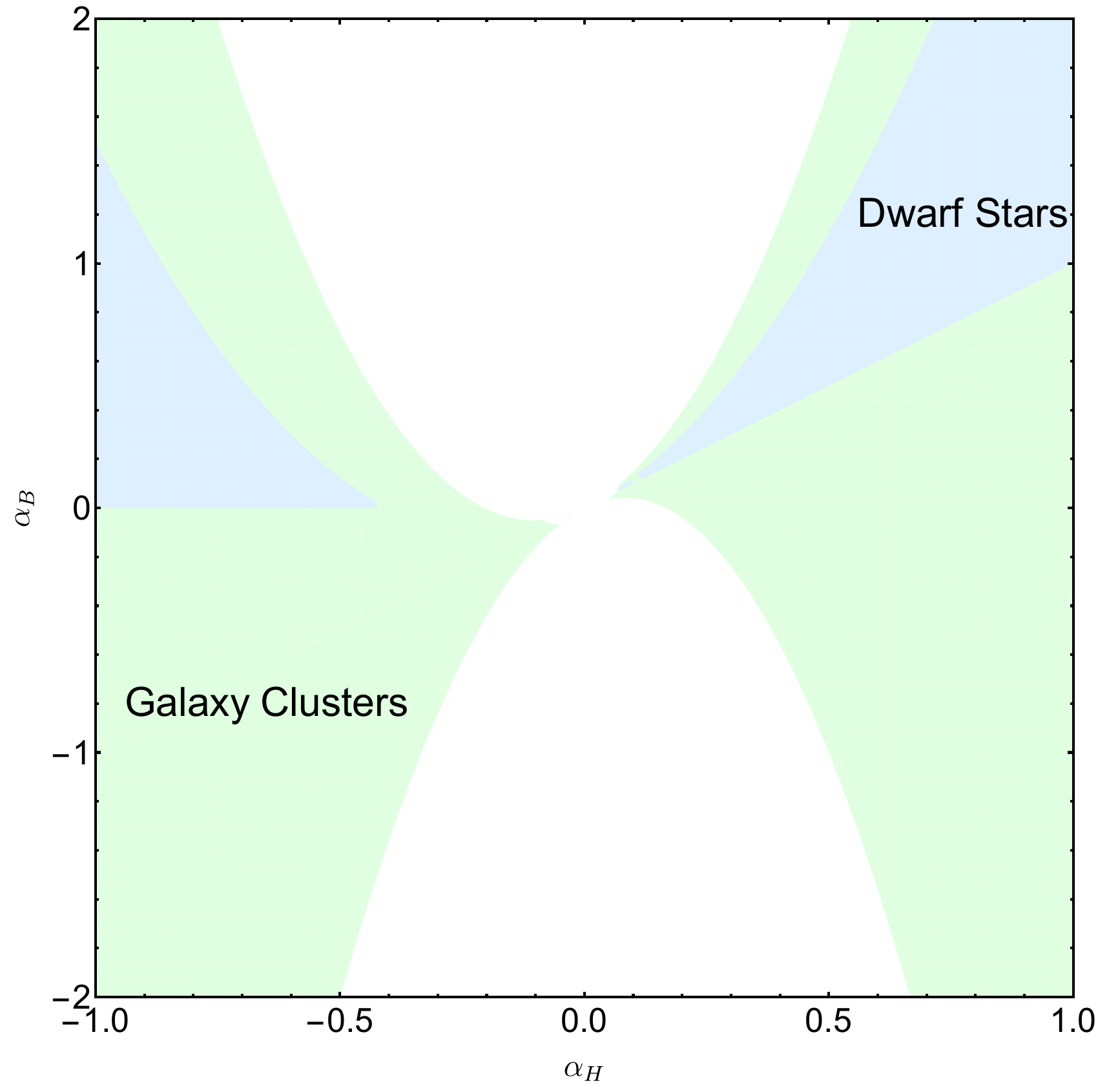}
    \caption{Bounds on the building blocks of dark energy $\alpha_B$ and $\alpha_H$ derived by translating bounds on $\Upsilon_1$ from dwarf stars \cite{Sakstein:2015zoa} (blue regions), and the bounds on $\Upsilon_1$ and $\Upsilon_2$ from galaxy clusters \cite{Sakstein:2016ggl} (green regions) via equations \eqref{eq:U1NEW} and \eqref{eq:U2NEW}.~Figure taken from \cite{Sakstein:2017xjx}.}   \label{fig:beyondHorndeskiLocal}
\end{figure}

An example is given in Figure \ref{fig:beyondHorndeskiLocal}, which shows exclusion regions in the $\alpha_B$--$\alpha_H$ plane.~This figure was created using constraints from the minimum mass for hydrogen burning in dwarf stars, which limits $\Upsilon_1<1.6$ \cite{Sakstein:2015zoa,Sakstein:2015aac}, and from comparing weak lensing and X-ray masses of galaxy clusters, yielding $\Upsilon_1=-0.11^{+0.93}_{-0.67}$ and $\Upsilon_2=-0.22^{+1.22}_{-1.19}$ \cite{Sakstein:2016ggl}.

Shortly after our analysis \cite{Sakstein:2017xjx}, the expressions for $\Upsilon_i$ in terms of the dark-energy building blocks were generalized to the full GLPV framework \cite{Dima:2017pwp,Langlois:2017dyl,Crisostomi:2017lbg}, allowing this first implication of GW170817/GRB170817A to be extended to a wider class of theories.~Continued improvements in small-scale probes will further tighten these bounds and may ultimately reveal deviations from GR.

\subsection{Quartic Galileon Theories Cannot Explain Dark Energy}

\begin{figure}
\centering
\includegraphics[width=0.5\textwidth]{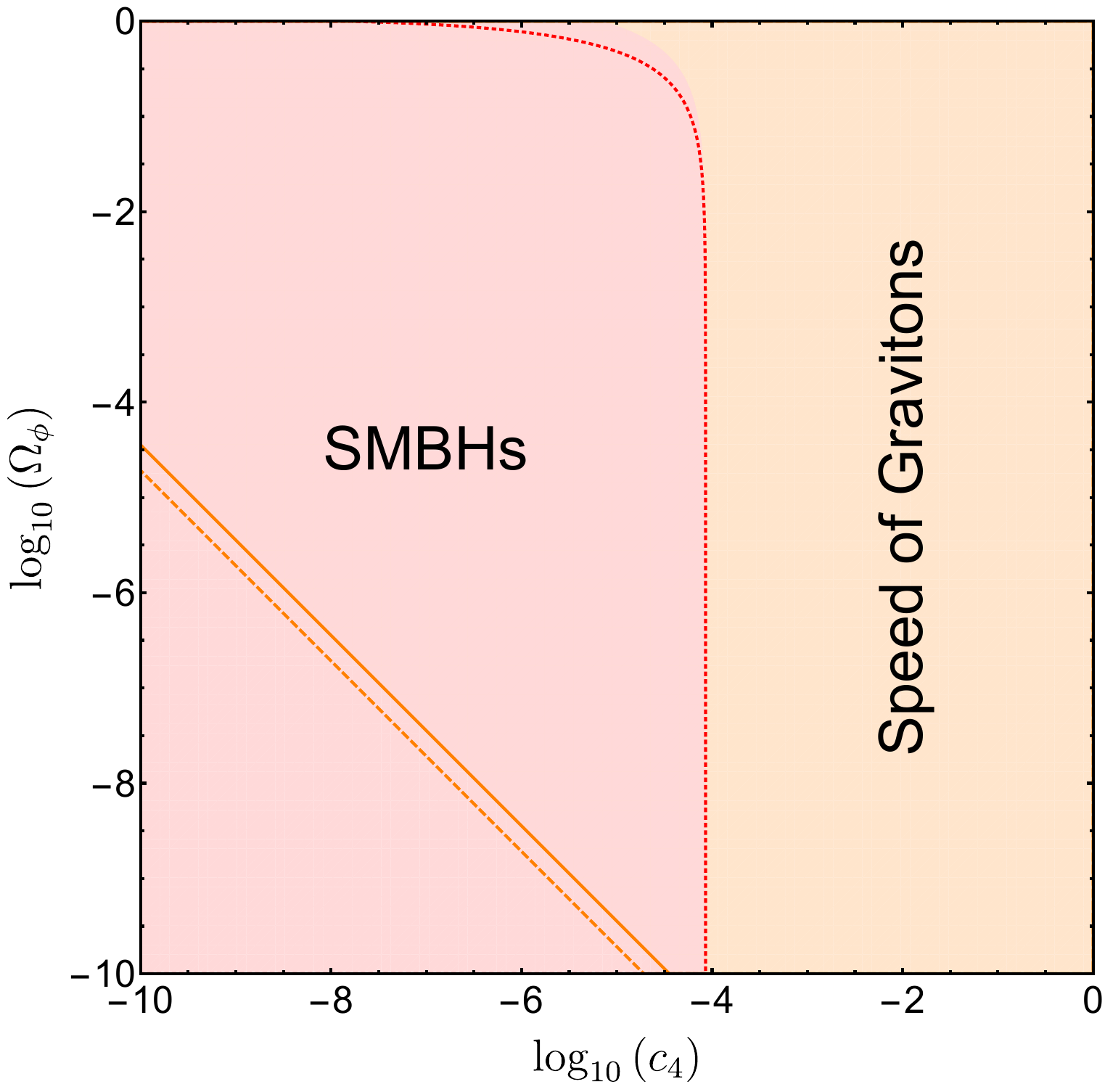}
\caption{Constraints in the $c_4$–$\Omega_\phi$ plane from the near equality of the speed of gravitons and photons (orange) and the lack of a supermassive black hole offset in M87 (red).~Shaded regions are excluded for $c_0=1$; dashed/dotted curves show equivalent limits for $c_0=3$.~Figure taken from \cite{Sakstein:2017xjx}.}
\label{fig:quartic}
\end{figure}

As mentioned above, \textit{covariant quartic galileons} \cite{Deffayet:2009wt} --- introduced in section~\ref{sec:VainshteinMechanism} --- were a benchmark for GLPV dark energy theories prior to GW170817/GRB170817A.~Their Lagrangian is
\begin{align}
\frac{\mathcal{L}}{\sqrt{-g}} &= X + G_4(\phi,X)R  
+ G_{4,\,X}\left[(\Box\phi)^2 - \nabla_\mu\nabla_\nu\phi\,\nabla^\mu\nabla^\nu\phi\right], \label{eq:quarticLagrangian} \\
G_4(\phi,X) &= \frac{M_{\rm Pl}^2}{2}\left(1 + 2c_0\frac{\phi}{M_{\rm Pl}}\right) + 2\frac{c_4}{\Lambda_4^6}X^2, \label{eq:quarticG4}
\end{align}
where $X=-(\partial\phi)^2/2$, $c_0$ controls the matter coupling, and $c_4$ the quartic interaction.~The speed of gravitational waves on FLRW backgrounds in this theory is \cite{Brax:2015dma}  
\begin{equation}
\left|{c_T^2-1}\right| = \left|\frac{4c_4x^2}{1-3c_4x^2}\right|, \quad \textrm{with}\quad  
x \equiv \frac{\dot{\phi}}{HM_{\rm Pl}},
\label{eq:quarticGWspeed}
\end{equation}
and the amount of dark energy relative to the total energy content of the universe is \cite{Appleby:2011aa}
\begin{equation}
\Omega_\phi = c_0 x^2 + \frac{x^2}{6} + \frac{15c_4x^2}{2}.
\label{eq:quarticOmega}
\end{equation}
The LIGO/Virgo–Fermi bound on $|c_T^2 - 1|$ therefore constrains a combination of $c_4$ and the cosmological parameters of the model.

On small scales, \cite{Sakstein:2017bws} used the absence of strong equivalence principle violations to bound $c_0$ and $c_4$.~In quartic Galileon gravity (and GLPV theories more generally), non-relativistic matter couples to the scalar field but black holes do not.~As a result, a galaxy’s supermassive black hole (SMBH) and its stars would fall at different rates in an external gravitational field, producing a measurable SMBH offset from the galactic center \cite{Hui:2012jb}.~Applying this idea to M87 in the Virgo cluster yields strong constraints on small $c_4$, complementary to the large $c_4$ bounds from gravitational wave speed measurements.

Figure~\ref{fig:quartic} shows the combined constraints in the $c_4$–$\Omega_\phi$ plane for representative $c_0$ values.~The gravitational wave speed bound rules out large $\Omega_\phi$ at high $c_4$, while the SMBH test excludes large $\Omega_\phi$ at low $c_4$.~Together, they imply $\Omega_\phi \ll 1$ for all allowed $c_4$, excluding the quartic Galileon as the  driver of the cosmic acceleration.~One can also apply the bound on the speed of light vs.~gravity to exclude more general GLPV theories \cite{Creminelli:2017sry,Ezquiaga:2017ekz,Baker:2017hug} with some caveats \cite{deRham:2018red,Copeland:2018yuh}.

\subsection{The Disformal Coupling to Photons is Strongly Constrained}

Beyond GLPV theories, a further quantity that is constrained by GW170817/GRB170817A is the disformal coupling of the scalar field to photons.~At leading order, and neglecting conformal couplings (which do not affect the speed of light or gravitational waves), the metric in equation~\eqref{eq:disformal} can be written as 
\begin{equation}
\tilde{g}^{(i)}_{\mu\nu} = g_{\mu\nu} + \frac{\partial_\mu\phi\,\partial_\nu\phi}{\mathcal{M}_i^4},
\label{eq:disformal_metric}
\end{equation}
where $\mathcal{M}_i$ is the disformal coupling scale and we have allowed for different couplings to different matter species $i$.~If all species share the same $\mathcal{M}_i$, the equivalence principle is preserved, but generically the photon and baryon couplings can differ \cite{vandeBruck:2016cnh}.~ 

In the simplest case with no disformal coupling to baryons, the photon speed is  
\begin{equation}
\frac{c_\gamma^2}{c^2} = 1 - \frac{\dot{\phi}^2}{\mathcal{M}_\gamma^4}.
\label{eq:photon_speed}
\end{equation}
The LIGO/Virgo–Fermi measurement of $|c_\gamma - c_{\rm GW}|/c \lesssim 6\times10^{-15}$ therefore implies  
\begin{equation}
\frac{\dot{\phi}^2}{\mathcal{M}_\gamma^4} \lesssim 6\times10^{-15}.
\label{eq:disformal_bound}
\end{equation}
For a cosmologically relevant scalar with $\dot{\phi} \sim H_0 M_{\rm Pl}$ \cite{Copeland:2006wr}, this bound translates to  
\begin{equation}
\mathcal{M}_\gamma \gtrsim 10\ \mathrm{MeV}.
\label{eq:Mgamma_bound}
\end{equation}
This limit exceeds that inferred from the absence of vacuum \v{C}erenkov radiation at LEP \cite{Hohensee:2009zk} and is comparable to bounds from solar energy-loss constraints via the Primakov process.~ 
It is weaker than the $\mathcal{M}_\gamma \gtrsim 100\ \mathrm{MeV}$ bound from horizontal branch star cooling \cite{Brax:2014vva}, but avoids the astrophysical uncertainties and systematics (e.g., metallicity) affecting stellar limits.~Imposing this constraint, the disformal coupling to photons is too small to influence the scalar’s cosmological evolution \cite{Sakstein:2015jca}.

\section{Outlook and Conclusions}

GW170817/GRB170817A was a landmark event that had profound implications for astrophysics and cosmology.~In this proceeding, we have summarized its role in constraining classes of scalar field dark energy, including GLPV theories and disformally coupled models, as first presented in our original paper \cite{Sakstein:2017xjx}.~Other dark energy theories outside of these classes were similarly constrained by an extensive body of work. Thus the observations of one  neutron star merger event had a profound implications for the theoretical landscape available to explain cosmic acceleration. 

In 2025, more than a quarter-century after the discovery of the cosmic acceleration, the phenomenon is more mysterious than ever.~Until last year, the cosmological constant --- where dark energy is static --- was the best-fitting model, but this changed in April 2024 when the Dark Energy Spectroscopic Instrument (DESI) reported strong evidence for evolving dark energy when combining baryon acoustic oscillation and type-Ia supernovae measurements with the cosmic microwave background \cite{DESI:2024mwx}.~Their next data release further increased the significance of this result \cite{DESI:2025zgx}.~Thus the search for viable fundamental models of dark energy has a new twist, and the theories that survive the stringent bounds from GW170817/GRB170817A form a  foundation for constructing such models.

\bibliography{ref}
\bibliographystyle{apsrev4-2}

\end{document}